\newif\ifdoublecolumn
\doublecolumntrue
\ifdoublecolumn
  \documentclass[journal,comsoc,9pt]{IEEEtran}
\else
  \documentclass[journal,12pt,onecolumn,draftclsnofoot]{IEEEtran}
  \usepackage{adjustbox}
\fi
\newif\ifanon
\anonfalse
\usepackage[T1]{fontenc}

\ifCLASSINFOpdf
\else
\fi
\usepackage{amsmath}
\usepackage[cmintegrals]{newtxmath}
\usepackage{array}

\usepackage[table,xcdraw]{xcolor}
\usepackage{color}
\usepackage{graphicx,pdfpages}
\usepackage{algorithm,algpseudocode,listings}
\usepackage[font=footnotesize]{caption}
\usepackage{tikz}
\usepackage{multirow}
\usepackage[plain]{fancyref}
\newcommand*{\fancyrefalglabelprefix}{alg}
\newcommand*{\frefalgname}{algorithm}
\newcommand*{\Frefalgname}{Algorithm}
\frefformat{plain}{\fancyrefalglabelprefix}{%
  \frefalgname\fancyrefdefaultspacing#1%
}%
\Frefformat{plain}{\fancyrefalglabelprefix}{%
  \Frefalgname\fancyrefdefaultspacing#1%
}%
\usepackage{hyperref}
\usepackage{subcaption}
\usepackage{pifont}
\newcommand{\cmark}{\ding{51}}
\newcommand{\xmark}{\ding{55}}
\newcommand{\smark}{\ding{229}}
\usepackage{etoolbox}
\newbool{darkrow}
\setbool{darkrow}{true}
\newcommand{\rowcol}{%
\ifbool{darkrow}%
{\rowcolor[HTML]{EFEFEF}\global\setbool{darkrow}{false}}%
{\rowcolor[HTML]{FFFFFF}\global\setbool{darkrow}{true}}%
}

\newcommand{\namelong}{Secure Open AP}
\newcommand{\nameshort}{SOAP}
\newcommand{\mc}[1]{\begin{tabular}[c]{@{}c@{}}#1\end{tabular}}
\newcolumntype{x}[1]{>{\centering\arraybackslash}m{#1}}

\begin{document}
\title{
Pre-shared Key Agreement for Secure Public Wi-Fi
}
\ifanon
  \author{Author 1, Author 2, Author 3}
\else
  \author{Seoksoeng Jeon, Chansu Yu and Young-Joo Suh%
  \thanks{
  }%
  \thanks{Seokseong Jeon is with Division of IT Convergence Engineering, POSTECH, Korea, Republic of (e-mail: s.jeon@postech.ac.kr)}%
  \thanks{Chansu Yu is with Department of Electrical and Computer Engineering, Cleveland state university, Ohio (e-mail: chansuyu@gmail.com)}%
  \thanks{Young-Joo Suh is with Department of Computer Science and Engineering, POSTECH, Korea, Republic of (e-mail: yjsuh@postech.ac.kr)}
  }
\fi
\markboth{Journal name}%
{Authors and title}
\maketitle

\begin{abstract}
This paper presents a novel pre-shared key (PSK) agreement scheme to establish a
secure connection between a Wi-Fi client and access point (AP) without prior
knowledge of a password.
The standard IEEE 802.11 security method, Robust Security Network Association,
widely known as Wi-Fi Protected Access (WPA) and WPA2, derives a
shared cryptographic key if and only if a user provides an identical password
which an AP possesses, causing of inconvenience of obtaining and entering the
password.
In this paper, a proposed scheme, {\it \namelong{} (\nameshort{})}, adopts two
public key algorithms, the elliptic curve Diffie-Hellman key exchange algorithm
(ECDH) and digital signature algorithm (ECDSA) to establish a secure connection
between a client and an AP without having prior knowledge of a password.
Implementation and experiment results demonstrate the viability of the proposed
scheme.
\end{abstract}

\section{Introduction}
\label{sec:intro}

With the wide distribution of mobile devices, Wi-Fi access points (APs) are
becoming more available in public areas.
They are mostly configured to use Open System Authentication (OSA) for clients'
convenience without having to enter a password.
However, OSA provides null authentication and no protection of data traffic
\cite{privacy} posing security and privacy threats.
On the other hand, a more recent IEEE 802.11 standard defines Robust Security
Network Association (RSNA) \cite{ieee80211-2016} for confidentiality and
integrity of data widely known as Wi-Fi Protected Access (WPA) and WPA2.
It provides secure communication at the cost of inconvenience of obtaining and
entering the password (pre-shared key or PSK).

In this paper, we propose {\it \namelong{} (\nameshort{})} which enables a
client and an AP to establish a secure connection without the prior knowledge of
a password.
Motivations of this work are three-fold:
First, it is desirable to make free and public Wi-Fi connections secure.
Second, captive portals, typically adopted in airports, coffee shops, etc,
appear to offer secure connections with a login page and legal notices.
However, they mostly use OSA and do not adequately warn that the communication
is not protected \cite{public-wifi-friend-or-foe}, \cite{ietf-capport}.
Third, authentication and security go hand in hand in Wi-Fi security algorithms.
SOAP separates the two and thus, allows an AP to require no authentication but
to provide message protection for each connected client.

To this end, \nameshort{} adopts two public key algorithms, a key agreement
algorithm and a digital signature algorithm \cite{menezes2012elliptic}.
First, without the loss of generality, the elliptic curve Diffie-Hellman key
exchange algorithm (ECDH) is used to agree on a PSK.
The elliptic curve digital signature algorithm (ECDSA) is
adopted to prevent a possible man-in-the-middle attack.
Along with it, \nameshort{} introduces a new information element (IE) and frame
format which conform to the standard and do not interfere
\nameshort{}-unaware legacy devices.
Experiment results show that \nameshort{} increases the connection delay by no
more than 22 percent
given the parameters used for the implementation.
This is mainly due to the increased management frame size and the additional
handshake messages but is not significant considering the benefit of clients'
security and convenience.

This paper is organized as follows:
\Fref{sec:primer} briefly explains authentication and security algorithms in
the IEEE 802.11 standard.
\Fref{sec:proposal} explains the proposed mechanism, \nameshort{}.
\Fref{sec:eval} examines its performance and finally we will conclude this study
in \fref{sec:concl}.

\section{Primer}
\label{sec:primer}

The IEEE 802.11 standard defines two classes of security algorithms, pre-RSNA
and RSNA algorithms as shown in \fref{tab:security-algs}.

\ifdoublecolumn
\begin{table}[]
\centering
\caption{IEEE 802.11 security algorithms}
\label{tab:security-algs}
\setlength\tabcolsep{4pt}
\begin{tabular}{|c|c|c|c|}
\hline
\begin{tabular}[c]{@{}c@{}}Security class\\ in IEEE 802.11\end{tabular}         & Authentication & \begin{tabular}[c]{@{}c@{}}Message\\ Protection\end{tabular} & Comment                                           \\ \hline
\multirow{2}{*}{Pre-RSNA}                                                            & OSA            & N/A                                                            & \begin{tabular}[c]{@{}c@{}}Convenient\\but no protection\end{tabular} \\ \cline{2-4} 
                                                                                     & SKA            & WEP                                                          & \begin{tabular}[c]{@{}c@{}}Some protection\\but found vulnerable\\(obsolete now)\end{tabular} \\ \hline
\multirow{2}{*}{\begin{tabular}[c]{@{}c@{}}RSNA\\(since\\802.11i-2004)\end{tabular}} & \multicolumn{2}{c|}{WPA/WPA2-EAP}                                             & \begin{tabular}[c]{@{}c@{}}Used in an enterprise\\environment with AS\end{tabular} \\ \cline{2-4} 
                                                                                     & \multicolumn{2}{c|}{WPA/WPA2-PSK}                                             & \begin{tabular}[c]{@{}c@{}}Used in open, public areas,\\Requires a client\\to enter a password\end{tabular} \\ \hline
\end{tabular}
\end{table}
\fi

Pre-RSNA includes Wired Equivalent Privacy (WEP) for message protection, and OSA
and Shared Key Authentication (SKA) for entity
authentication.
While SKA can be used if WEP has been selected, both are obsolete
due to their vulnerabilities and inefficiencies.
An AP in an OSA (\fref{fig:osa}) mode allows any client to connect to it without
verification of its legitimacy.

\ifdoublecolumn

\begin{figure}
  \centering
  \includegraphics[width=0.45\textwidth]{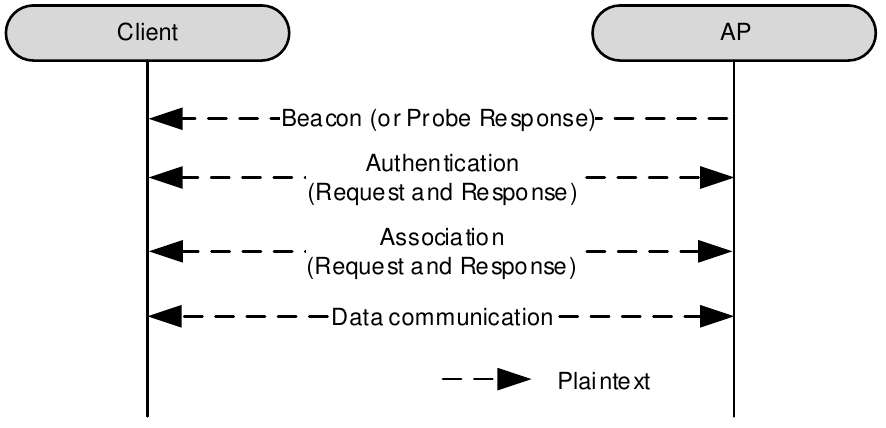}
  \caption{
    Open System Authentication procedure,
    which is null authentication and does not provide data protection
  }
  \label{fig:osa}
\end{figure}

\else

\begin{table}
\begin{adjustbox}{center}
  \begin{tabular}{>{\centering\arraybackslash}b{12cm}>{\centering\arraybackslash}p{8cm}}
\centering
\caption{IEEE 802.11 security algorithms}
\label{tab:security-algs}
\setlength\tabcolsep{4pt}
\begin{tabular}{|c|c|c|c|}
\hline
\begin{tabular}[c]{@{}c@{}}Security class\\ in IEEE 802.11\end{tabular}         & Authentication & \begin{tabular}[c]{@{}c@{}}Message\\ Protection\end{tabular} & Comment                                           \\ \hline
\multirow{2}{*}{Pre-RSNA}                                                            & OSA            & N/A                                                            & \begin{tabular}[c]{@{}c@{}}Convenient but no protection\end{tabular} \\ \cline{2-4} 
                                                                                     & SKA            & WEP                                                          & \begin{tabular}[c]{@{}c@{}}Some protection but found vulnerable\\(obsolete now)\end{tabular} \\ \hline
\multirow{2}{*}{\begin{tabular}[c]{@{}c@{}}RSNA\\(since\\802.11i-2004)\end{tabular}} & \multicolumn{2}{c|}{WPA/WPA2-EAP}                                             & \begin{tabular}[c]{@{}c@{}}Used in an enterprise environment with AS\end{tabular} \\ \cline{2-4} 
                                                                                     & \multicolumn{2}{c|}{WPA/WPA2-PSK}                                             & \begin{tabular}[c]{@{}c@{}}Used in open, public areas,\\Requires a client to enter a password\end{tabular} \\ \hline
\end{tabular}
    &
      \centering
      \includegraphics[width=0.45\textwidth]{figures/osa.pdf}
      \captionof{figure}{
        Open System Authentication procedure,
        which is null authentication and does not provide data protection
      }
      \label{fig:osa}
  \end{tabular}
  \end{adjustbox}
\end{table}

\fi

\begin{figure*}
  \centering
  \begin{minipage}[t]{0.49\linewidth}
    \begin{tikzpicture}[remember picture]
      \node[anchor=south west,inner sep=0](imageA)
        {\includegraphics[width=1\linewidth]{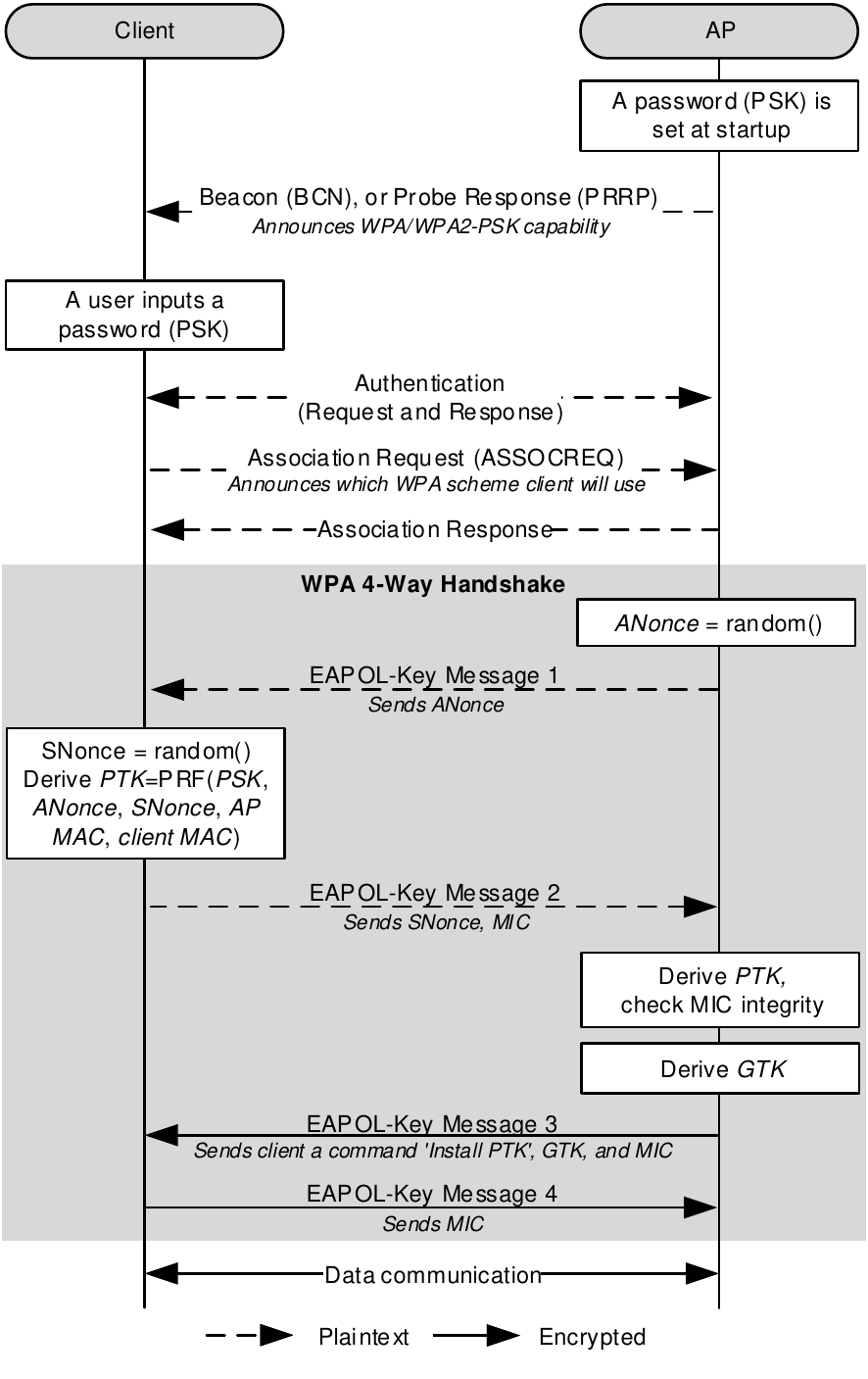}};
      \begin{scope}
        [shift=(imageA.south west),x={(imageA.south east)},y={(imageA.north west)}]
        \node[coordinate] (A) at (0.94,0.125) {};
        \node[coordinate] (B) at (1.023,0.125) {};
      \end{scope}
    \end{tikzpicture}
    \caption{
      The WPA-PSK procedure.
      (An AP sets a PSK at the configuration time and a client should enter an
      identical PSK to authenticate and derive a PTK.)
    }
    \label{fig:wpa-key-exchange}
  \end{minipage}
  \begin{minipage}[t]{0.49\linewidth}
    \begin{tikzpicture}[remember picture]
      \node[anchor=south west,inner sep=0](imageB)
        {\includegraphics[width=1\linewidth]{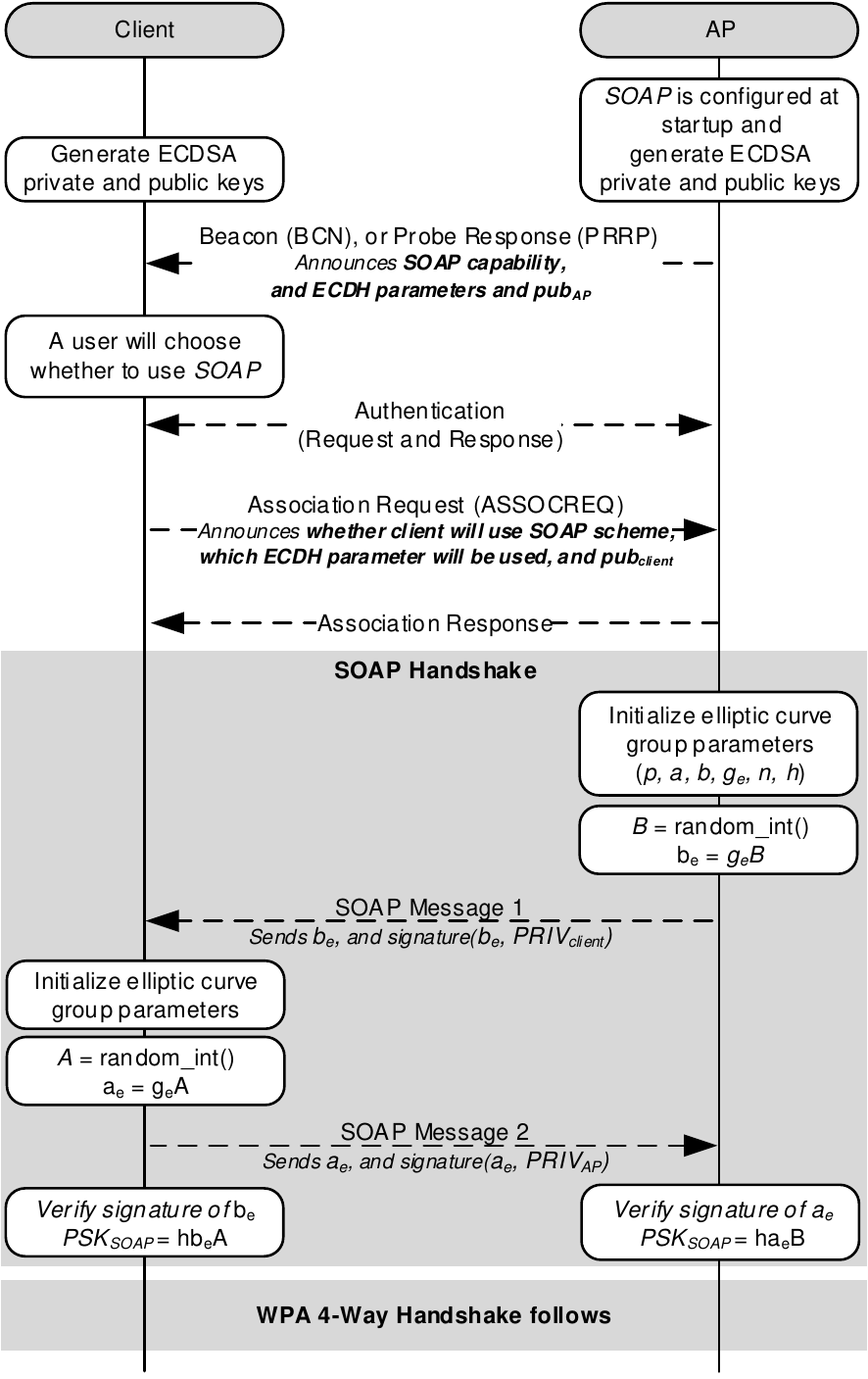}};
      \begin{scope}
        [shift=(imageB.south west),x={(imageB.south east)},y={(imageB.north west)}]
        \node[coordinate] (C) at (-0.023,0.025) {};
        \node[coordinate] (D) at (0.10,0.05) {};
      \end{scope}
    \end{tikzpicture}
    \caption{
      The \nameshort{} procedure.
      (Contrary to the existing WPA-PSK in \fref{fig:wpa-key-exchange}, a
      PSK is agreed on using the \nameshort{} Handshake without a user input.)
    }
    \label{fig:soap-key-exchange}
  \end{minipage}
  \begin{tikzpicture}[remember picture, overlay]
    \draw (A) edge[out=-90,in=180,->,line width=0.25mm] (D);
  \end{tikzpicture}
\end{figure*}

In RSNA, WPA can establish a secure connection between
a client and an AP if and
only if a client provides a valid user's membership information (WPA-EAP mode),
or if a client provides a correct password, PSK (WPA-PSK mode).
The former is used in, for example, companies or universities
using an authentication server (AS) such as RADIUS.
The latter is used in personal or open, small-scale public areas.
WPA-PSK establishes a connection between a client and an AP and derives a shared
cryptographic key (pairwise transient key or PTK) using an identical password,
PSK.
This is accomplished via the 4-Way Handshake as illustrated in
\fref{fig:wpa-key-exchange}.

Prior to the 4-Way Handshake, a client and an AP negotiate WPA parameters by
exchanging a Probe Response (PRRP), a Beacon (BCN), and an Association Request
(ASSOCREQ) frame.
Note that the WPA-PSK procedure in \fref{fig:wpa-key-exchange} begins with null
authentication (OSA) but it additionally provides a stronger authentication
during the 4-Way Handshake by proving knowledge of a shared
password.
More specifically, the 4-Way Handshake is triggered by the AP, which uses
Extensible Authentication Protocol over LAN (EAPOL) \cite{IEEE802.1X}.
With a PSK configured at startup on the AP side and given by a user input on the
client side, a PTK is derived from a PSK, SNonce, and ANonce on both sides.
The identity is assessed by checking the MIC, authenticating the client.
After the 4-Way Handshake, messages between the client and the AP are conveyed
as encrypted.
As described above, a PTK cannot be derived unless a client and an AP have a
common priori knowledge.

\section{\namelong{}}
\label{sec:proposal}

This paper aims for a user to be able to conveniently connect to an AP in a
WPA-PSK mode without having to enter a password.
\nameshort{} uses two public key algorithms, ECDH and ECDSA, prior to the 4-Way
Handshake to agree on a PSK on both sides so that the handshake can use it for
the 4-Way Handshake as shown in \fref{fig:soap-key-exchange}.
\Fref{alg:dh} summarizes the \nameshort{} Handshake.


\begin{algorithm}[h]
  \small
  \caption{\nameshort{} Handshake}
  \label{alg:dh}
  \begin{algorithmic}[1]
    \Statex \textbf{Inputs}
    \Statex \ \ $\{pub, PRIV\}_{\{client,AP\}}$: ECDSA public/private keys of a client and an
                                                 AP
    \Statex \textbf{Output}: a shared secret key $PSK_{\nameshort{}}$
    \State Exchange ECDH groups and ECDSA public keys via PRRP/BCN, and ASSOCREQ
    \State Negotiate an ECDH group $G$ (\fref{alg:ie})
    \State Initialize an ECDH generator (primitive element) $g_e$
    \State Generate ECDH private keys
\ifdoublecolumn
    \Statex $\left(\begin{array}{l}
              Client: A = random\_int()\\
              AP: B = random\_int()
            \end{array}\right., 1 \le A,\ B \le p-1$
\else
    \Statex $\begin{array}{ccc}
              Client: A = random\_int() &
              AP: B = random\_int(), &
              1 \le A,\ B \le p-1
            \end{array} $
\fi
    \State Generate ECDH public keys
    \Statex $\begin{array}{cc}
              Client: a_e = g_eA & AP: b_e = g_eB
            \end{array}$
    \State Exchange $a_{e}$ and $b_{e}$ with their signature via \nameshort{} Message 1/2
    \State Agree on a shared secret
\ifdoublecolumn
    \Statex $\left(\begin{array}{l}Client: 
                     PSK_{\nameshort{}} = hb_eA
                   \\
           AP: 
                   PSK_{\nameshort{}} = ha_eB
                 \end{array}\right.=hg_eAB$
\else
    \Statex $\begin{array}{ccc}
                     PSK_{\nameshort{},client} = hb_eA &
                     =hg_eAB &
                     = ha_eB = PSK_{\nameshort{},AP}
                 \end{array}$
\fi
  \end{algorithmic}
\end{algorithm}

\subsection{\nameshort{} Information Element and Elliptic Curve Negotiation}
\label{sec:ie}

In \fref{fig:soap-key-exchange}, 
a client and an AP first negotiate whether they will use \nameshort{}, and if
so, which elliptic curve they will use.
Such information is conveyed via a newly introduced IE\footnotemark{}, \nameshort{} IE, which
is included in the frame body of a PRRP, a BCN and an ASSOCREQ
\footnotetext{
  Note that BCN, PRRP, or ASSOCREQ are management frame subtypes.
  The frame body of a management frame consists of IEs, format of which is a 1
  octet Element ID (EID) field, a 1 octet Length field, and a variable length
  element-specific Information field.
  Predefined elements are SSID (EID 0), Supported Rates (EID 1), TIM (EID 5),
  etc.
}
%
and is defined as in \fref{fig:soap-ie}.
Element ID is set to 251 which is a reserved value and is not used in the
current standard.
Group count $m$ represents the number of available elliptic curve groups and
Group list is a list of 1-octet long integer identifiers representing elliptic
curve groups.

\begin{figure}[h]
  \centering
  \ifdoublecolumn
\centering
\small
\setlength\tabcolsep{4pt}
\begin{tabular}{c|c|c|c|c|c|c|}
\cline{2-7}
       & \multirow{2}{*}{\begin{tabular}[c]{@{}c@{}}Element\\ ID\end{tabular}} & \multirow{2}{*}{Length} & \multicolumn{4}{c|}{Information}                                                                                                                                                                                                                   \\ \cline{4-7} 
       &                                                                       &                         & \begin{tabular}[c]{@{}c@{}}Group\\ count\\ ($m$)\end{tabular} & \begin{tabular}[c]{@{}c@{}}Group\\ list\end{tabular} & \begin{tabular}[c]{@{}c@{}}ECDSA\\ key\\ size ($s$)\end{tabular} & \begin{tabular}[c]{@{}c@{}}ECDSA\\ public\\ key\end{tabular} \\ \cline{2-7} 
Octets & 1                                                                     & 1                       & 1                                                           & $m$                                                    & 1                                                              & $s$                                                            \\ \cline{2-7} 
\end{tabular}
  \else
\centering
\setlength\tabcolsep{4pt}
\small
\begin{tabular}{c|c|c|c|c|c|c|}
\cline{2-7}
       & \multirow{2}{*}{\begin{tabular}[c]{@{}c@{}}Element\\ ID\end{tabular}} & \multirow{2}{*}{Length} & \multicolumn{4}{c|}{Information}                                                                                                                                                                                                                   \\ \cline{4-7} 
       &                                                                       &                         & \begin{tabular}[c]{@{}c@{}}Group count ($m$)\end{tabular} & \begin{tabular}[c]{@{}c@{}}Group list\end{tabular} & \begin{tabular}[c]{@{}c@{}}ECDSA key size ($s$)\end{tabular} & \begin{tabular}[c]{@{}c@{}}ECDSA public key\end{tabular} \\ \cline{2-7} 
Octets & 1                                                                     & 1                       & 1                                                           & $m$                                                    & 1                                                              & $s$                                                            \\ \cline{2-7} 
\end{tabular}
  \fi
  \caption{\nameshort{} IE}
  \label{fig:soap-ie}
\end{figure}

The AP first announces a set of available elliptic curve groups,
$\textbf{G}_{AP}$, via a \nameshort{} IE included in a PRRP/BCN.
Upon receiving a PRRP/BCN, the client having its own supported set of groups,
$\textbf{G}_{client}$, (i) computes a common set
$\textbf{G}_\cap=\textbf{G}_{AP}\cap\textbf{G}_{client}$, (ii)
selects $G$ among the set with the largest key size , i.e., the highest security
level, and (iii) sends it back to the AP via a \nameshort{} IE included in an
ASSOCREQ.

ECDH group negotiation can be summarized as \fref{alg:ie}.

\begin{algorithm}[h]
  \small
  \caption{ECDH group negotiation}
  \label{alg:ie}
  \begin{algorithmic}[1]
    \Statex \textbf{Inputs}
    \Statex $\ \ \textbf{G}_{\{client,AP\}}$: A set of ECDH groups available on each side
    \Statex \textbf{Output}
    \Statex $\ \ G$: A negotiated ECDH group
    \State An AP sends a PRRP/BCN to a client with $\textbf{G}_{AP}$
    \State $Client: \textbf{G}_\cap = \textbf{G}_{AP} \cap \textbf{G}_{client}$
    \If{$\textbf{G}_\cap = \varnothing$} \Comment {WPA-PSK mode}
      \State The client requires a user to enter a password
      \State The client sends an ASSOCREQ to the AP without $G$
    \Else \Comment {\nameshort{} mode}
      \State The client selects $G$ with the largest key size among $\textbf{G}_\cap$
      \State The client sends ASSOCREQ to with $G$ to the AP
    \EndIf
  \end{algorithmic}
\end{algorithm}
\subsection{\nameshort{} Handshake and Message Formats}
\label{sec:handshake-message}

After ECDH group negotiation is completed, the client and the AP first
initializes ECDH machines with the negotiated group $G$ and generates their own
private keys $A, B$ and public keys $a_e, b_e$, respectively.
The client and the AP exchange their ECDH public keys via \nameshort{} Message
1 and 2.
The AP sends its ECDH public key with its signature via a \nameshort{} Message 1
to the client.
Receiving the \nameshort{} Message 1, the client sends $a_e$ and its signature via a
\nameshort{} Message 2 back to the AP.
After exchanging and verifying a \nameshort{} Message from each other, the
client and the AP must be able to agree on a $PSK_{\nameshort{}}$.
And the AP triggers the 4-Way Handshake and both use $PSK_{\nameshort{}}$ instead of
$PSK$ during the handshake.

Frame body of \nameshort{} Message 1 and 2 is defined as shown in
\fref{fig:soap-message}.
\begin{figure}[h]
  \centering
  \ifdoublecolumn
\centering
\small
\setlength\tabcolsep{4pt}
\begin{tabular}{c|c|c|c|c|c|}
\cline{2-6}
& \multicolumn{3}{c|}{EAPOL header}& \multicolumn{2}{c|}{Packet body} \\
\cline{2-6}
& \mc{Protocol\\version} & \mc{Packet\\type} & \mc{Packet\\body\\length} & \mc{ECDH\\public\\key} & \mc{ECDSA\\signature} \\ \cline{2-6} 
Octets & 1                                                          & 1                                                     & 2                                                            & variable & variable  \\ \cline{2-6} 
\end{tabular}

  \else
\centering
\setlength\tabcolsep{4pt}
\small
\begin{tabular}{c|c|c|c|c|c|}
\cline{2-6}
& \multicolumn{3}{c|}{EAPOL header}& \multicolumn{2}{c|}{Packet body} \\
\cline{2-6}
& \mc{Protocol version} & \mc{Packet type} & \mc{Packet body length} & \mc{ECDH public key} & \mc{ECDSA signature} \\ \cline{2-6} 
Octets & 1                                                          & 1                                                     & 2                                                            & variable & variable  \\ \cline{2-6} 
\end{tabular}

  \fi
  \caption{
    Frame body of a \nameshort{} Message.
  }
  \label{fig:soap-message}
\end{figure}
Like an EAPOL-Key frame used in the 4-Way Handshake, a link layer control (LLC)
header is prepended to \nameshort{} Message 1 and 2 so that they are routed in a
different reception and processing path than a normal data frame.
In other words, a \nameshort{} Message is a variant of an EAPOL-Key frame.
Protocol version and packet type are currently set to \lstinline{0xff} which
is a reserved value and is not used in the current IEEE 802.1X standard.
The EAPOL header is followed by ECDH public key and its signature, whose
lengths are determined by the previous negotiation process.




Note that messages for the \nameshort{} Handshake are signed with the ECDSA 
private keys as mentioned earlier.
The \nameshort{} IE in \fref{fig:soap-ie} contains an ECDSA public key of a
sender (the client or the AP) to be used in the \nameshort{} Handshake.
It is used to sign the \nameshort{} messages.
SHA-256 is used for digestion.
The client and the AP can verify the authenticity of a counterpart with the
signature.
\section{Evaluation}
\label{sec:eval}

\subsection{Feasibility and Applicability}

\subsubsection{Implementation}

We implemented \nameshort{} on a Linux laptop.
Kernel versions of 3.13 and 4.4, and Atheros AR9565 Wi-Fi NIC supporting IEEE
802.11b/g/n are used for evaluation.
Most Linux distributions use \lstinline{wpa_supplicant} as a backend daemon for
a WPA client.
We modified it to implement \nameshort{} on a client side.
Note that \lstinline{wpa_supplicant} operates in user space, and can be used on
other operating systems, and thus exhibits great applicability.
%
For an evaluation purpose, \nameshort{} is implemented using \lstinline{hostapd}
on an AP side.
Both a client and an AP support 224-bit (28-octet) long key for both ECDH and
ECDSA.
The implementation is publicly accessible via
\ifanon\cite{SOAP-anon}\else\cite{SOAP}\fi
.

\subsubsection{Coexistence with legacy devices}

First, according to the standard, an IE with an unknown element ID in a
management frame shall be discarded silently and it does not produce any side
effect (BCN, PRRP, or ASSOCREQ implementing \nameshort{}).
Second, while a \nameshort{} Message is an EAPOL-Key frame variant with an
invalid EAPOL header, it shall not be generated unless a \nameshort{} IE is
exchanged prior to the \nameshort{} Handshake.
Even if it is transmitted by misbehavior, a \nameshort{}-unaware device shall
discard it silently, too.
Note that a \nameshort{}-aware client and AP also can establish a connection
using a legacy WPA-PSK mode.

\subsection{Security Analysis}
\label{sec:security-analysis}

Since \nameshort{} relies on the solid foundations of ECDH and ECDSA, a
\nameshort{}-capable client and AP are able to securely agree on a PSK.
This section enumerates potential threats to \nameshort{} and presents the
corresponding security analysis as shown in \fref{tab:security-analysis}.

In the table, the vulnerability due to message injection and masquerading are inherited from
OSA.
A client may connect with a rogue AP and is allowed a DoS attack and privacy
leakage.
We expect it can be resolved by obtaining ECDSA keys being managed globally for
trustful APs, which is analogous to HTTPS certificates.
Another vulnerability of connection hijacking can be caused by injecting a
disassociation frame before \nameshort{} handshake, which applies to WPA in the
same manner.
Note that it cannot hijack a connection (session), and only can
disconnect a client from an AP.
This can be resolved if management frames are also signed and verified with
ECDSA keys.


\begin{table*}[]
  \small
\centering
\caption{Security analysis of \nameshort{}}
\label{tab:security-analysis}
\begin{tabular}{>{\arraybackslash}p{0.25\linewidth}>{\arraybackslash}p{0.75\linewidth}}
  \hline
  \rowcol
  \textbf{Threat or criterion} & \textbf{Security analysis (\cmark{}: Secure, \xmark{}: Vulnerable, \smark{}: Possible solution)} \\
  \hline
  \rowcol
  Ephemeral $PSK_{SOAP}$ &
  \hangindent=1em
  \hangafter=1
  \cmark{} Ephemeral $PSK_{\nameshort}$ is obtained at every new \nameshort{} handshake \\
  \rowcol
  Elliptic curve key size &
  \hangindent=1em
  \hangafter=1
  \cmark{} 224-bit elliptic curve is acceptable until 2022 or later \cite{Barker_2016, BSI} \\
  \rowcol
  Acitve/passive eavesdropping &
  \hangindent=1em
  \hangafter=1
  \cmark{} $PSK_{\nameshort}$ remains secret owing to confidentiality property of ECDH \\
  \rowcol
  Message replaying &
  \hangindent=1em
  \hangafter=1
  \cmark{} Properly implemented \nameshort{} state machine can prevent replay attack by
  discarding replayed frames \\
  \rowcol
  Message deletion and interception &
  \hangindent=1em
  \hangafter=1
  \cmark{} It can cause DoS but requires techniques such as jamming which apply to all
  other Wi-Fi communication \\
  \rowcol
  Message injection &
  \hangindent=1em
  \hangafter=1
  \xmark{} It can cause DoS if PRRQ, PRRP and BCN containing an incorrect ECDSA public key are injected \newline
  \smark{} It can be filtered by blacklisting a device with repetitive ECDSA verification failure \\
  \rowcol
  Masquerading (MAC/SSID spoofing) &
  \hangindent=1em
  \hangafter=1
  \xmark{} Vulnerable as OSA if a client and an AP accept PRRQ, PRRP and BCN frames from an attacker \newline
  \smark{} Once correct ECDSA public keys are exchanged, masquerading has no effect \\
  \rowcol
  Connection hijacking &
  \hangindent=1em
  \hangafter=1
  \xmark{} An attacker can disconnect a client from an AP by sending a disassociation frame, causing DoS \newline
  \smark{} But it cannot hijack due to the attacker has a different ECDSA public key \newline
  \smark{} Disassociation can be prevented if management frames are signed and verified with ECDSA \\
  \hline
\end{tabular}
\end{table*}

\subsection{Network Performance Analysis}

\subsubsection{Connection establishment delay}
\label{sec:delay}

\begin{figure}[h]
  \centering
  \includegraphics[width=0.4\textwidth]{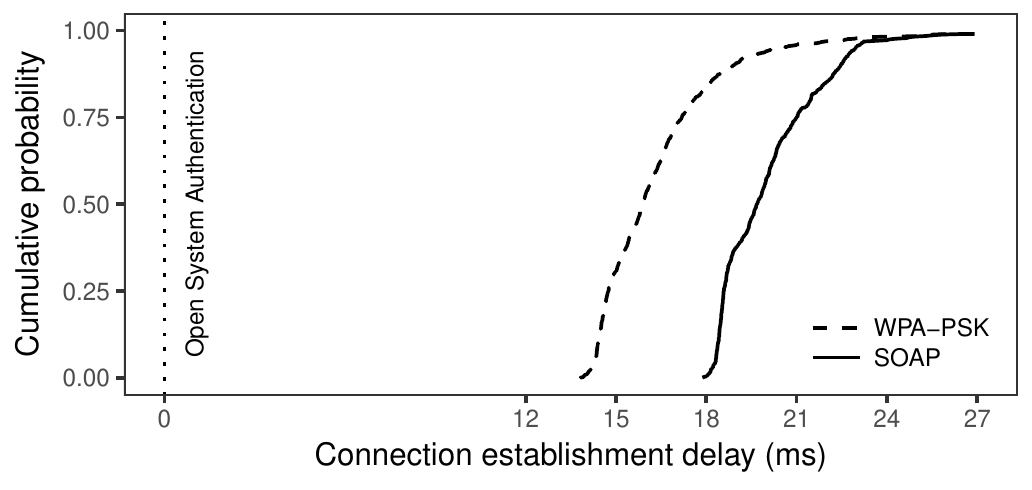}
  \caption{
    Cumulative probability of connection establishment delay.
    (Mean delay is 16.44 ms and 20.09 ms without and with \nameshort{} scheme,
    respectively.
    Mean difference is 3.65 ms and it corresponds to approximately 22 percent
    additional time delay.)
  }
  \label{fig:delay}
\end{figure}

An extra delay due to the \nameshort{} Handshake is inevitable.
We measure the time taken to establish a secure connection between a client and
an AP.
The delay is measured on an AP side, and is defined by a time interval from
when an Association Response frame is sent and to when an EAPOL-Key Message 4
is sent.
\lstinline{hostapd} is modified
to measure time of each event in microsecond precision.
It is measured 1,000 times, plotted and compared in \fref{fig:delay}.
The mean extra delay caused by the \nameshort{} Handshake is 3.65 ms, which
corresponds to approximately 22 percent higher delay compared to the 4-Way Handshake (WPA-PSK).

Note that this delay is only involved when a client and an AP attempt to
establish a connection.
Hence the impact of additional delay due to \nameshort{} is trivial.

\subsubsection{Network Overhead}

\begin{figure}[h]
  \centering
  \ifdoublecolumn
  \begin{minipage}[t]{5.1cm}
  \else
  \begin{minipage}[t]{0.5\textwidth}
  \fi
    \centering
    \includegraphics[width=5.1cm]{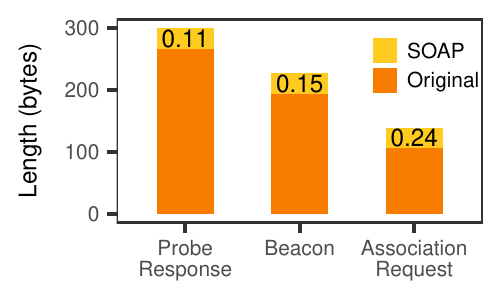}
    \subcaption{Management frame overhead due to \nameshort{} IE}
    \label{fig:eval-overhead}
  \end{minipage}%
  \begin{minipage}[t]{3.4cm}
    \centering
    \includegraphics[width=3.4cm]{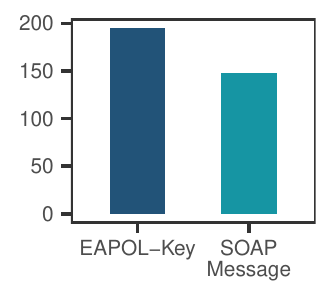}
    \subcaption{\nameshort{} Message size}
    \label{fig:eval-message}
  \end{minipage}
  \caption{Frame length evaluation}
  \label{fig:eval-frame}
\end{figure}

Overheads of a \nameshort{} IE and a \nameshort{} Message are
measured and compared in \fref{fig:eval-frame}.
It is assumed that a client and an AP support one common ECDH group and use a
224-bit elliptic curve as discussed earlier, thus $m=1$ and $s=28$ in
\fref{fig:soap-ie}, and a \nameshort{} IE occupies 33 bytes,
where ECDSA Public key field takes up a major portion
(28 octets; 85 percent).
It corresponds to up to 24 percent of overhead as shown in
\fref{fig:eval-overhead}.
Adding another ECDH group to the IE only takes up 1 more octet.

Size of a \nameshort{} Message is calculated as the sum of ECDH key
size and ECDSA signature size, which is double of ECDSA key size.
Assuming ECDH and ECDSA use an 224-bit elliptic curve, a \nameshort{} Message is
148 bytes long including a MAC header, while an EAPOL-Key frame is 195 bytes
long as shown in \fref{fig:eval-message}.

\section{Conclusion}
\label{sec:concl}

Motivated by the need of open but secure Wi-Fi communication, this paper
proposes \nameshort{}, which renders a client and an AP to securely agree on a
PSK using two public key algorithms as an extension of WPA-PSK.
The proposed scheme adopts a key agreement algorithm, ECDH and a digital
signature algorithm, ECDSA for that purpose.

We expect that \nameshort{} can secure a public Wi-Fi network without
introducing inconvenience for a provider and a consumer, and also that it can be
used to construct a secure IoT network with automated process.
We further think \nameshort{} can be extended to use a certificate instead of
a simple public key algorithm and signature to manage and assess a trusted
public AP in a global scale against a rogue AP.



\bibliographystyle{IEEEtran}
\bibliography{IEEEabrv,includes/ref}
\end{document}